\documentclass{aastex63}

\usepackage{amsmath,upgreek} 

\newcommand\pennstate{Department of Astronomy \& Astrophysics and \\
Center for Exoplanets and Habitable Worlds and \\
Penn State Extraterrestrial Intelligence Center \\
525 Davey Laboratory \\
The Pennsylvania State University \\
University Park, PA, 16802, USA}

\begin{document}
\title{A Framework for Relative Biosignature Yields from Future Direct Imaging Missions}

\author[0000-0003-3989-5545]{Noah W.\ Tuchow}
\affil{\pennstate}

\author[0000-0001-6160-5888]{Jason T.\ Wright}
\affil{\pennstate}

\begin{abstract}

Future exoplanet direct imaging missions, such as HabEx and LUVOIR, will select target stars to maximize the number of Earth-like exoplanets that can have their atmospheric compositions characterized. Because one of these missions' aims is to detect biosignatures, they should also consider the expected biosignature yield of planets around these stars.

In this work, we develop a method of computing relative biosignature yields among potential target stars, given a model of habitability and biosignature genesis, and using a star's habitability history. As an illustration and first application of this method, we use MESA stellar models to calculate the time evolution of the habitable zone, and examine three simple models for biosignature genesis to calculate the relative biosignature yield for different stars.  

We find that the relative merits of K stars versus F stars depend sensitively on model choice. In particular,  use of the present-day habitable zone as a proxy for biosignature detectability favors young, luminous stars lacking the potential for long-term habitability. Biosignature yields are also sensitive to whether life can arise on \textit{Cold Start} exoplanets that enter the habitable zone after formation, an open question deserving of more attention.

Using the case study of biosignature yields calculated for $\uptheta$ Cygni and 55 Cancri, we find that robust mission design and target selection for HabEx and LUVOIR depends on: choosing a specific model of biosignature appearance with time; the terrestrial planet occurrence rate as a function of orbital separation; precise knowledge of stellar properties; and accurate stellar evolutionary histories.

\end{abstract}


\section{Introduction}

One of the main goals in the study of planets around other stars is to characterize the atmospheric compositions of Earth-like planets\deleted{,} and determine whether they host biosignatures. Based on the detection or non-detection of unambiguous biosignatures in the spectra of planetary atmospheres, we aim to constrain the occurrence rates of life on other planets. For future missions, one would like to gain a sense for the yield of biosignatures that one would expect to find. In this paper, we will introduce a framework to characterize which stars have the greatest chances of hosting biosignatures on their planets. We will start by investigating how the biosignature yield changes as a function of stellar age, and then we will consider effect\added{s} of changing both stellar mass and age. We will then discuss how different assumptions for planet occurrence rate and the emergence of biosignatures affect which types of stars are preferred in terms of biosignature yield and consider which yield metrics would be most useful for future mission planning.

\subsection{Direct Imaging}
To characterize the atmospheric compositions of Earth-like planets, the optimal detection technique would be to directly image them. However, designing an instrument capable of blocking out a star's light to allow observations of the very faint light reflected off a planet's surface is a major engineering challenge. The direct imaging method of exoplanet detection favors larger planets at high separations from their stars, and all planets found to date using direct imaging have been longer period, giant planets. To extend the capacities of direct imaging to be able to observe Earth sized planets in the habitable zones of their stars, one requires a large next generation telescope and a sophisticated coronagraph with a sufficiently small inner working angle. 

For the 2020 Astrophysics Decadal Survey, NASA has solicited four Large Mission Concept studies, two of which aim to directly image habitable Earth-sized planets. The Large UV/Optical/Infrared Surveyor (LUVOIR) is a concept for a large telescope capable of observing wavelengths ranging from the UV to the infrared \citep{LUVOIR_final_report}. LUVOIR has two potential architectures: \replaced{a}{the} 15 meter diameter LUVOIR-A and 8 meter LUVOIR-B. Both of these architectures use an internal coronagraph to block out a star's light, but they differ with regards to estimated cost and the design of the coronagraph. The other main direct imaging mission concept under consideration is the Habitable Exoplanet Observatory (HabEx) \citep{HabEX_Final_report}. While the HabEx telescope has a  4 meter diameter, smaller than the LUVOIR telescope, it makes up for it by having a 52m starshade external to the telescope. In addition to the starshade, HabEx has an alternative vector vortex coronagraph on the primary spacecraft and a UV spectrograph.

For future direct imaging missions, it is important to gain a sense for the estimated number of Earth-like planets for which we will be able to obtain spectra.  \citet{Stark2019} have calculated the estimated yield of Earth-like exoplanets for different mission architectures. They suggest that internal coronagraph-based missions may have yields that increase steadily with telescope diameter, while starshade missions appear to have diminishing returns past a certain aperture size. 
Such yield estimates are dependent on a variety of factors. For instance, to calculate the number of Earth-sized planets one expects to image, one needs to have prior knowledge of $\eta_\Earth$, the occurrence rate of Earth-like planets. Given the large uncertainties in current estimates of $\eta_\Earth$, exoplanet yield estimates are similarly uncertain.
Furthermore, yield calculations need to consider the exposure times required to obtain enough light to construct a quality spectrum \citep{Stark2015}. Exposure times will depend on the brightness of the host star and planet, the amount of exozodiacal background light in the system, and the required signal to noise ratio.  One must also consider the time required to maneuver the telescope to view each target star, especially in the case of a starshade mission, and the fuel requirements of the spacecraft \citep{Stark2016}. 

To maximize the yield of habitable Earth sized exoplanets or biosignatures, future missions will need to select an optimal sample of target stars. Such a list of target stars has not been finalized yet due in part to the fact that many or most of the planets that these missions will characterize have likely not been discovered. \deleted{yet.}
Preliminary lists are limited to bright nearby stars because of the faintness of the reflected light from exoplanets and the angular separations required to resolve a planet separately from its star.
The stars which a mission chooses to observe depends on the observational strategy employed. If one utilizes a strategy of ``statistical comparative planetology'' as described by \citet{Bean2017}, one aims to obtain observations of a few key properties of a large number of planets to test our models of planetary habitability. Alternatively, a complementary strategy is to focus on a smaller number of exoplanets most likely to host biosignatures and thoroughly characterize them.
Regardless of the approach one takes, one must design missions around a coherent strategy that considers which stars are most likely to host habitable planets.
In this paper, we will develop a framework to rank stars according to their potential to host planets with observable biosignatures. This will serve as a basis for future studies aiming to assess the biosignature yield of direct imaging missions.

\subsection{Habitable Zones}
A mission to search for biosignatures should have the ability to image planets in the habitable zones of their stars. While the habitable zone is not a specific or complete description of planetary habitability, it has proved to be an invaluable tool in the search for habitable exoplanets because it provides a fiducial target for the discovery of truly Earth-like planets. Defined as as the region around a star where a rocky planet could support liquid water on its surface, given a $\text{CO}_2$-$\text{H}_2\text{O}$-$\text{N}_2$ atmosphere, the conventional habitable zone provides astronomers with a general location for where to search for Earth-like planets \citep{Maunder1913,Huang1959,Kasting1993hz,Lorenz2020}. Given the flux constraints for the inner and outer edges, it is straightforward to calculate the habitable zone around any star of known luminosity and temperature. 

Formulations of the habitable zone vary based on assumptions about planetary atmospheric physics and composition. Some models consider only $\text{CO}_2$ and $\text{H}_2\text{O}$ to be the prominent greenhouse gasses in planetary atmospheres, while others extend the definition of the habitable zone, including dry planets with low $\text{H}_2\text{O}$ concentrations, and considering alternative sources of greenhouse heating such as $\text{H}_2$ atmospheres \citep{Kasting2014,Kopparapu2013,Abe2011,Pierrehumbert2011}. Given the large variability in the habitable zone boundaries, future missions aim to observe the planets around the general location of the habitable zone, and determine empirically where habitable planets occur.

While the habitable zone gives a range of distances where an Earth-like planet may be habitable, it is not the only location where habitable planets may be found. For instance, in our own solar system, moons of giant planets such as Europa and Titan may host habitable environments. However, given the challenges in assessing the habitability of such locations in our own solar system, assessing the habitability of analogous environments around other stars may be insurmountably difficult, and biosignatures in these locations may not be detectable.   

\subsection{Habitable Histories}

Knowledge that a planet exists in its star's habitable zone does not necessarily mean that it is habitable. Rather than depending on the bulk properties of a planet, such as mass, radius, and orbital period, habitability is dependent on the climatic and geological history of a planet \citep{Lenardic2016}. Planets exhibiting the same bulk properties, may be vastly different in terms of their potential for habitability. One of the the key influences on the climatic history of a planet is that stars do not stay constant in time. Therefore, as the luminosity and effective temperature of a star evolve in time, so too does its habitable zone.

To characterize the habitability of an exoplanet we need the following properties of its host star. The star's luminosity and effective temperature have a fairly straightforward effect on a planet's habitability in that they define the current day \textit{instantaneous habitable zone} (IHZ). The mass and age of a star are also required to assess the habitability of its planets. These have a less obvious effect on habitability, determining the evolutionary history of a star and how its luminosity and temperature, and thus its habitable zone, evolve in time \citep{Gallet2019}. Whereas the luminosity and effective temperature of a star can be determined via direct observations, the mass and age of a star require a stellar model to be fit to the observed properties of a star. Masses and ages are therefore harder to obtain and typically have much larger uncertainties. In addition to these stellar properties, other factors such as metallicity, helium abundance and rotation rates play an important role in the evolution of stars and their habitable zones \citep{Danchi2013,Valle2014}. Metallicity from a scaled solar abundance pattern may not be enough to describe habitable zone evolution, as stellar composition and the abundances of specific chemical species may have a significant effect on stellar evolutionary tracks \citep{Truitt2015,Truitt2017}.
It is clear that in order to tell whether a planet is habitable and how long it has been habitable, one needs to be familiar with many of the properties of its host star.

For a planet to be a potential host for biosignatures, it must have maintained habitable conditions long enough for life to originate and develop. 
Planets that currently reside in the habitable zone, but haven't been there for long are sub-optimal targets for direct imaging missions, as life wouldn't have time to develop.
In the case of the Earth, life is believed to have evolved very early in the Earth's lifetime \citep{Schopf2006}. 
However, it wasn't until 2.4 - 2.1 Gyr ago, during the Great Oxidation Event (GOE), that oxygen from photosynthesis appeared as a potentially detectable biosignature \citep{Kasting1993earth,Lyons2014}. Even after the GOE, oxygen levels in the Earth's atmosphere may have been comparatively much lower than today, lasting up until the end of the Proterozoic eon, roughly 600 Myr ago \citep{Crockford2018}. It is unclear whether or not oxygen would have been observable to a hypothetical observer prior to this period. If one assumes that life on other planets may evolve on similar timescales to life on Earth, to detect observable biosignatures, one would need to look at an Earth analog that has been in the habitable zone for at least 2 Gyr \citep{Truitt2015}. Thus, the present day habitable zone alone is insufficient information to determine a planet's likelihood to host biosignatures. 

The concept of a \textit{continuously habitable zone} (CHZ) has been used to define a region around a star where planets remain habitable over an extended period of time. There is no one definition for a continuously habitable zone, but definitions usually range between having the planet remain in the habitable zone for 2 Gyr, to 4 Gyr, to the star's entire main sequence lifetime \citep{Truitt2015,Buccino2006,Valle2014}. Past studies have also considered the amount of time which a planet spends in the habitable zone over the course of its star's main sequence evolution, which has been referred to by different names by different studies, such as the habitable zone lifetime or duration of transit of the habitable zone \citep{Rushby2013,Danchi2013,Valle2014,Waltham2017}. 

\section{Metrics for long-term habitability}
\label{metrics_section}

\begin{figure}[t!]
    \centering
    \plotone{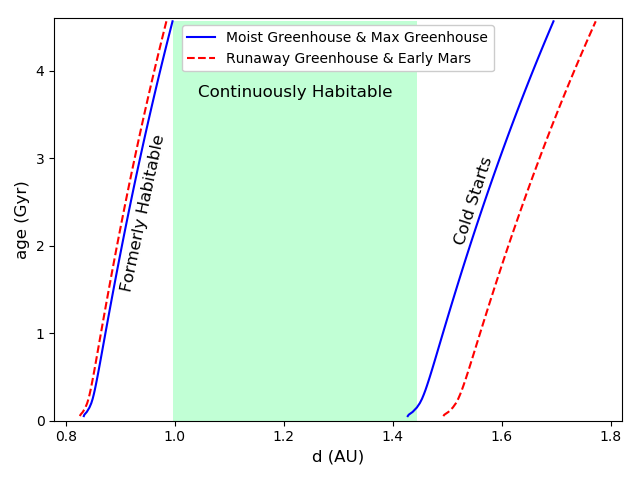}
    \caption{Habitable zone around a sun-like star as it evolves in time to the current day. The green region denotes the continuously habitable zone (CHZ) as defined in this study, starting after 0.2 Gyr. Habitable zones \added{are} calculated using the \citet{Kopparapu2013} formulation. Stellar models used are described in Appendix \ref{Stellar_models}.}
    \label{Sun_hz}
\end{figure}

\subsection{Habitable Durations}
For the purpose of life detection, it bears little consequence if a planet will spend a large amount of time in the habitable zone {\it in the future}; only the character and duration of its {\it past} habitability matters.
Therefore, for this study, we use a definition of the CHZ as the region around a star that has remained in the star's habitable zone from 0.2 Gyr after formation, roughly the age for which we have the earliest evidence of Earth supporting a solid surface and liquid water \citep{Wilde2001}, up until present. Past studies, such as that of \citet{Hamano2013}, suggest that analogous terrestrial planets should have surfaces that solidify after around 100 Myr at the most, so the use of the Earth-based timescale for a solid surface is a conservative upper estimate for the beginning of habitability. 

Figure \ref{Sun_hz} shows how the habitable zone of a sun-like star changes as a function of time. Here we use the \citet{Kopparapu2013} formulation of the habitable zone and show the evolution of both the conservative and optimistic definitions of the inner and outer edges. The CHZ as it is defined in this paper is shown in green, using the conservative definition of the habitable zone, where the inner and outer edges are defined by the moist greenhouse limit and maximum greenhouse heating respectively. The stellar model shown here is a MESA model described in Appendix \ref{Stellar_models}, using approximately solar properties, with a mass of $M=1.00 M_\odot$, initial helium and metal mass fractions of $Y_i=0.28$ and $Z_i=0.02$, and an age of $4.57$ Gyr \citep{Bahcall1995}.

As we are concerned with the emergence of biosignatures on habitable exoplanets, the amount of time that a planet has spent in the habitable zone prior to the current day is of principal interest. We next define the {\it habitable duration}, $\tau$, as the time that a hypothetical planet at a fixed orbital radius has spent in the habitable zone before present. 
Habitable duration is similar to the ``habitable lifetime'' of \citet{Waltham2017} and the ``duration of transit'' of the habitable zone of  \citet{Danchi2013,Valle2014}.  These other definitions are typically used in a more theoretical context, and consider the amount of time a planet would spend in the habitable zone over the entire main sequence and post-main sequence evolution of its host star, whereas our habitable duration considers only the past history of a given star.

\begin{figure}
    \centering
    \includegraphics[width=5in]{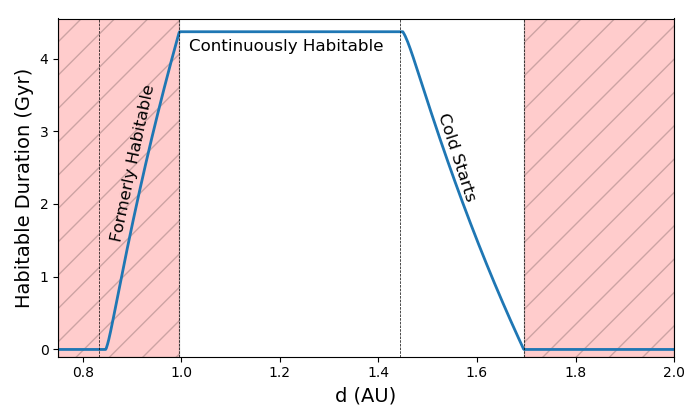}
    \caption{Habitable duration, $\tau$, as a function of orbital separation for a sun-like star of age 4.57 Gyr. Red regions are outside the present-day, instantaneous habitable zone.}
    \label{past_hab}
\end{figure}

In Figure \ref{past_hab}, we plot the habitable duration as a function of orbital distance from a star. 
This plot is constructed using the same model for stellar evolution of a sun-like star seen in Figure \ref{Sun_hz}, and the habitable duration can be thought of as the span on the time axis  bounded by the inner and outer habitable zone boundary curves and the present age and minimum age for habitability.
There are 4 dotted vertical lines in Figure \ref{past_hab}. The first and third lines from the left represent the inner and outer boundaries of the habitable zone when the star was on the zero age main sequence, while the second and fourth line from the left give the habitable zone boundaries at the present time. The red region in this plot represents regions which are currently outside of the habitable zone, and the plateau in the center represents the CHZ. The portion of the curve in the red region between the first and second line represents planets that were formerly in the habitable zone but have since become too hot.
On the right of Figure \ref{past_hab}, the downward sloping region of the curve between the third and fourth lines are planets that were outside of the habitable zone, but have entered it as the star's luminosity increases and the habitable zone has expanded. These are the so called {\it Cold Start} planets addressed in \citet{Kasting1993hz}.\footnote{This should not be confused with the term ``cold start'' used in the context of giant planet formation, an unrelated concept.}

\subsection{Cold Starts}
\label{cold_starts_section}
The habitability of Cold Start planets, originating outside the habitable zone, but entering it later in their lifetime, remains dubious, or uncertain at best. Different studies disagree as to whether these planets could thaw and become habitable. The high albedo of a globally glaciated planet would require a large stellar flux to thaw, perhaps greatly exceeding that of the outer habitable zone boundary \citep{Kasting1993hz}. \citet{Yang2017} suggest that flux required to thaw a fully glaciated planet may result in a sudden shift to a moist greenhouse atmosphere, resulting in significant water loss and hydrogen escape. Alternatively, \citet{Wolf2017} suggest that if a Cold Start planet were to support  present day Earth levels of CO$_2$, there would be a narrow range of stellar fluxes where it could support a habitable climate for an extended period of time. However, it remains unclear whether the these globally glaciated Cold Start planets would have sufficient outgassing to allow large enough concentrations of greenhouse gases in their atmospheres. Cold Start planets are an area which warrants future research, and yet are often naively assumed to be habitable. With only knowledge of the instantaneous habitable zone of a star, one is unable to tell whether a given planet has been continuously habitable, or if it belongs to this class of Cold Starts. Future direct-imaging missions will need to choose to target or avoid Cold Start planets, so we will perform our analysis for both cases. 

Cold Start planets become increasingly numerous as a star evolves. During their main sequence evolution, most stars gradually increase in luminosity, causing their habitable zones to move outwards. In Fig \ref{past_hab_lifetime_age}, we show how the habitable duration curve changes as a function of time for the sun-like star modelled in Figs \ref{Sun_hz} and \ref{past_hab}. This figure shows a variation of the habitable duration curve, where regions outside of the current habitable zone are set to zero. One can observe that, as the star ages, the plateau region of the curves\deleted{,} representing the CHZ gets progressively narrower as the inner edge recedes. On the other hand, the downward sloping region representing Cold Starts occupies an increasing portion of the habitable zone as the star ages and more planets enter the habitable zone.

\subsubsection{Uncertainty in Stellar Ages}
\label{age_uncertainty}
Perhaps the most important stellar property for determining the long-term habitability of planets is the stellar age. We have seen that the habitable duration and the inner and outer boundaries of the CHZ are heavily dependent on the age of the star. Unfortunately stellar ages are often difficult to constrain, with degenerate masses and ages corresponding to similar effective temperatures and luminosities. 
\deleted{It is not uncommon for stellar ages to have uncertainties on the order of many Gyr, amounting to factors of a few.}

\added{While masses can be constrained with a decent degree of certainty, ages are among the more difficult stellar properties to determine. This is evident in the median uncertainties provided in the \textit{Gaia-Kepler} Stellar Properties Catalog. While stellar masses have a median uncertainty of only 7\%, the median age uncertainty is significantly higher at 56\% \citep{Berger2020a}. One should note that many of the stars in the Kepler dataset may be poorly studied. These stars lack precise measurements of their observable properties, and they may possess unseen binary companions, contributing to the uncertainty in ages. Furthermore, low mass stars such as M dwarfs evolve very slowly over the course of billions years, so it is very difficult to constrain their ages to any precision. 

Typically the mass, age, and metallicity of a star are obtained via fitting stellar models to the observed properties of the star. This can be done using a precomputed grid of models, as in isochrone fitting, or by stellar model inversion, typically running an optimizer and evaluating the stellar model at each step. The later option allows for more finely sampled points and thus more accurate stellar output properties, but it is also much more computationally intensive. Using asteroseismic constraints in addition to spectroscopic constraints allows model fitting to obtain much more precise ages, but these methods can only be used for the few stars which possess sufficiently accurate asteroseismic data \citep{Soderblom2010,Bellinger2019}. An alternative means to determine stellar ages is via gyrochronology, where ages are inferred based on empirical relations between stellar rotation period and ages \citep{Barnes2003}. Such a method is only applicable for stars with measurable rotation periods, which are non-trivial to obtain, but for these stars one can potentially obtain model independent constraints on stellar ages. Pasts studies such as \citet{Claytor2020}, have been able to use gyrochronology methods to obtain ages for cool dwarf stars to a median uncertainty of 14\%. While these age measurements inferred from stellar rotation are more precise than those of isochrone fitting, in the case of low mass stars they may be biased based on factors such as the use of standard solar abundance patterns and uncertainty in the physics of stellar angular momentum transport. Given that stars vary in terms of their measurable properties and have different uncertainties in those measurements, every target star in a sample will likely have a different uncertainty in its inferred age.}

This difficulty in determining stellar ages motivated the Bayesian analysis of \citet{Truitt2020}, which considers the case where a star's age is unknown. For a planet at a given distance from its star, their study calculates the probability that the planet spends more than 2 Gyr in the habitable zone for cases where either the stellar mass or metallicity are known. Their study did not consider the obstacles for habitability for Cold Start planets, and did not incorporate the known present-day luminosity of a star, so for a star of solar mass and metallicity, their metrics prefer planets at Mars-like distances (because Mars will spend a lot of its {\it future} in the habitable zone, while an Earth-like planet will not). 
In our study, we will instead rank stars according to their potential to support planets with long habitability histories at any distance, without prior knowledge about whether they host planets, but incorporating our (often quite precise) knowledge of the instantaneous habitable zone. \added{Modelling the evolutionary tracks for stars with more precise ages from model inversion,} we will determine how different assumptions about the occurrence rates of planets and the emergence of biosignatures change which sorts of stars should be preferred by direct imaging missions.

\subsection{Relative Biosignature Yield}
\label{B_formulation}

To assess the potential for a star to host planets with biosignatures, we need a metric for biosignature yield that incorporates what we know about a star's long-term habitability even in the absence of knowledge about its particular planetary system. We define the {\it relative biosignature yield}, $B$, for a given star as
\begin{equation}
        B = \iint H(a,t) \Gamma(a,R_p)\,da\,dR_p
        \label{metric}
\end{equation}
where $H(a,t)$ is the probability that a planet at orbital separation $a$ from a star of age $t$ hosts detectable biosignatures, and where $\Gamma(a,R_p) = \frac{\partial^2 N}{\partial a \partial R_p}$ is the distribution of the number of planets per star in semimajor axis $a$ and planetary radius $R_p$.
\added{In this integral, $a$ is integrated over all values, while $R_p$ is integrated over the range of values for habitable, rocky exoplanets. Since $H(a,t)$ is independent of $R_p$, this equivalent to using a marginalized form of $\Gamma$ for the distribution of habitable exoplanets in $a$.}
$B$ is then an \replaced{expectation value}{expected yield} for the number of biosignatures \deleted{expected} on the planets orbiting the star.
 
 Obviously, we do not know the absolute value of $H$, but we can make reasonable inferences as to its functional form, which can guide which stars are the best targets. \added{Using the same form of $H$, one can compare \textit{relative} biosignature yields between target stars without knowledge of the exact probabilities of biosignature genesis.} \deleted{For instance, if we expect that a planet within the habitable zone is more likely to host detectable biosignatures than one outside of it, then we can compare the {\it relative} biosignature yields of stars with many habitable zone planets and those without such planets, without knowledge of the absolute value of $H$.
 }
 Past studies have made many different assumptions for $H(a,t)$ (sometimes tacitly). One may, for example, naively assume that all planets in the instantaneous habitable zone (IHZ) have an equal chance of hosting biosignatures. Studies, such as \citet{Stark2019}, have calculated expected yields of Earth-like planets in the IHZ for direct imaging missions, but it should be noted that estimating the yield of Earth-like planets in the habitable zone and estimating the yield of biosignatures are not the same. If a mission to study biosignatures were to synonymize the two, then this would be equivalent to assuming:
 
 \begin{equation}
 H(a,t) = 
     \begin{cases}
     \text{constant}, & \text{if } a \text{ in IHZ} \\
     0,               & \text{otherwise}
     \end{cases}
 \end{equation}
 
 \noindent where, again, the absolute value of the constant is not needed for comparing $B$ between two stars.
 
 Another choice could be to consider all planets that remain habitable beyond a given amount of time, say 2 Gyr, to have the same chance of hosting biosignatures. This would have the form of
 \begin{equation}
 H(a,t) = 
 \begin{cases}
 0,           & \text{if } \tau(a,t) < 2 \text{ Gyr} \\
 \text{constant}, & \text{otherwise}
 \end{cases}
 \end{equation}
where $\tau(a,t)$ is the habitable duration. This is the common ``2 Gyr CHZ'' as used by, for instance, \citet{Truitt2020}.

One could also suppose that biosignatures have a constant rate of developing per unit time, $b$, on any planet within the habitable zone. Then we can frame the biosignature probability as 
\begin{equation}
 H(a,t) = 1-e^{-b\tau(a,t)}   
\end{equation}
\noindent where again $\tau$ is the habitable duration of the planet. If $b$ is very small (i.e. there is a low chance of biosignatures arising per unit time) then $H(a,t)$ can be safely approximated as 

\begin{equation}
\label{LinearH}
    H(a,t) \propto \tau(a,t). 
\end{equation}

A more general (and perhaps more realistic) case would be where there is both a constant probability of biosignature creation per unit time, $b$, and a constant probability of destruction per time, $x$. This would take the form of 
 \begin{equation}
     H(a,t) = \frac{b}{b+x} \left[1 - e^{-(b+x) \tau(a,t)} \right]
 \end{equation}
 In the case of small $x+b$, $H(a,t)$ can again be approximated as $H(a,t) \propto \tau(a,t)$, as in Equation~\ref{LinearH}. 
 
 For the cases where $H(a,t)$ depends on $\tau(a,t)$, there are two formulations of $\tau$ that one can use. If only planets in the CHZ are considered habitable, then $\tau$ is set to zero outside the CHZ. Alternatively, if both Cold Start planets and planets in the CHZ are assumed to be habitable, then $\tau$ is only set to zero outside the IHZ.

In this work, we will consider three of these models for $H(a,t)$: the IHZ model, and two models with constant probability of biosignature creation with time, one assuming Cold Starts are valid and one using only the CHZ. The value of $b$ in these models is, of course, unknown (just as the constant value of H(a,t) is unknown in the IHZ model), but by restricting ourselves to the {\it relative} yields expected among stars, in the small $b$ case, we need only consider the time dependence of H(a,t) and leave $b$ as a parameter to be determined from observations.

The distribution of exoplanets in semimajor axis, $\Gamma(a,R_p)$, is also a function that is not well known for the regions of parameter space that we are interested in, but it is a quantity that future missions hope to determine. \added{With the success of missions such as \textit{Kepler}, we now have a large population of planets of known periods and radii, which can be used to empirically constrain the planetary distribution function \citep{Youdin2011}. Distribution functions typically take the form of power laws or split power laws in period and radius. However, we lack a sufficiently large sample of Earth-sized planets in the habitable zones of sun-like stars, requiring either extrapolation or inferences based on limited data \citep{Petigura2013,Burke2015}. Furthermore, it appears that $\Gamma$ is dependent on stellar properties such as effective temperature and metallicity, though this relationship is not well understood \citep[e.g.][]{Garrett2018}.}

For the purposes of this paper, our goal is not to choose the best \replaced{value}{form} of $\Gamma$ to use, but to determine what effects it has on the values of the metrics. Therefore we will consider two contrasting estimates for $\Gamma(a,R_p)$:  one that is uniform in $a$ and one that is uniform in $\ln(a)$. One should note that the true form of $\Gamma$ is almost certainly dependent on more than just $a$ and $R_p$\deleted{. In particular, the planet occurrence rate likely depends on stellar properties such as the star mass and metallicity. However, the dependence of $\Gamma$ as a function of these properties is not well understood. Therefore, in this paper we will only consider the two previously mentioned forms}\added{, but in this study we will only consider the two forms} of $\Gamma$ that are independent of stellar properties with the purpose of illustrating how different forms of $\Gamma$ affect expected biosignature yields.

\subsection{For Known Planet Hosts}

The aforementioned metrics primarily apply to the case where a target star is not known to host planets, for instance in the early stages of mission design before planets are discovered. These metrics then serve to estimate how likely a given star is to host long-term habitability and biosignatures based on planet occurrence rates and the evolution of the star's habitable zone. However, some future mission designs prioritize nearby stars known to host exoplanets. As such, it would be useful to have a variation of our biosignature yield metric for target selection among known planet hosts.

Rather than using the distribution $\Gamma(a,R_p)$ for planet occurrence rates, we can instead use a function $\phi(a,R_p)$ for the measured posterior distributions of a known planet's semimajor axis and radius. If the detailed distribution of planetary distances and radii aren't known, then $\phi$ could be easily be approximated as a 2D Gaussian with a diagonal covariance matrix using the squared uncertainties in $a$ and $R_p$. With the substitution of $\phi$ instead of $\Gamma$, the biosignature yield can be reformulated as
\begin{equation}
    B = \sum_{i=1}^{N} \iint H(a,t) \phi_i(a,R_p)\,da\,dR_p
\end{equation}
where here we consider the sum over $N$ planets in the system, where $\phi_i(a,R_p)$ is the distribution in parameters for each planet.

Note that comparing the biosignature yields between known planet hosts and star\deleted{'}s without known planets depends heavily on the extent to which a future mission prioritizes planet host stars. Further consideration is also needed for the case where only some of a star's planets are known. For example systems such as 55 Cancri, which are known to host giant planets, but may contain undetected Earth-sized planets in their habitable zones, the probability of planet occurrence is presumably influenced in complex ways that lie outside the scope of this work. 
Because the purpose of this paper is to illustrate how to calculate biosignature yields, and because most direct-imaging target planets haven't been discovered yet, we will restrict our analysis to the formulation of $B$ in Section \ref{B_formulation}.

\begin{figure}
    \centering
    \includegraphics[width=4in]{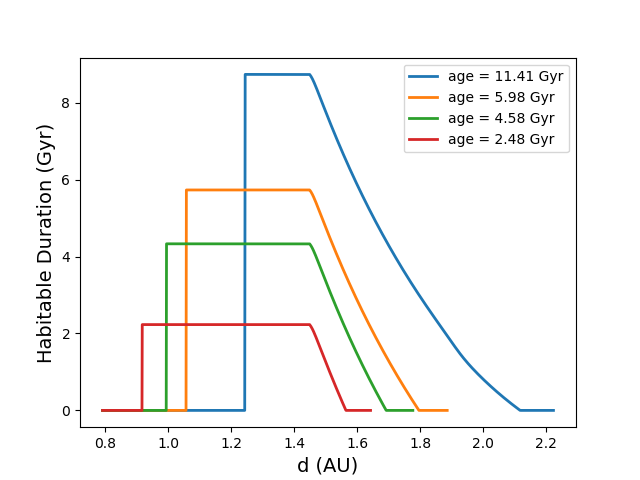}
    \caption{Habitable duration, $\tau$ as a function of orbital distance for a $1 M_\odot$ star at different ages, but with formerly-habitable planets (i.e.\ those outside of the present-day instantaneous habitable zone) set to zero. }
    \label{past_hab_lifetime_age}
\end{figure}

\section{Variability of Biosignature Yield Metrics}
\label{B_variability}

Using the formulation of our metrics $B$ in \replaced{equation}{Equation}  \ref{metric}, we consider six different metrics used to infer relative biosignature yields corresponding to the 3 functional forms of $H(a,t)$ and 2 forms of $\Gamma(a,R_p)$ described in Section \ref{B_formulation}. 
With these six metrics, we would like observe how $B$  compares for different kinds of stars. 
We will first see how these metrics change as a function of time for a given stellar model with fixed properties, and then we will observe the variation of the metrics over a wide range of stellar masses and ages. 

\begin{figure}
    \centering
    \includegraphics[width=6in]{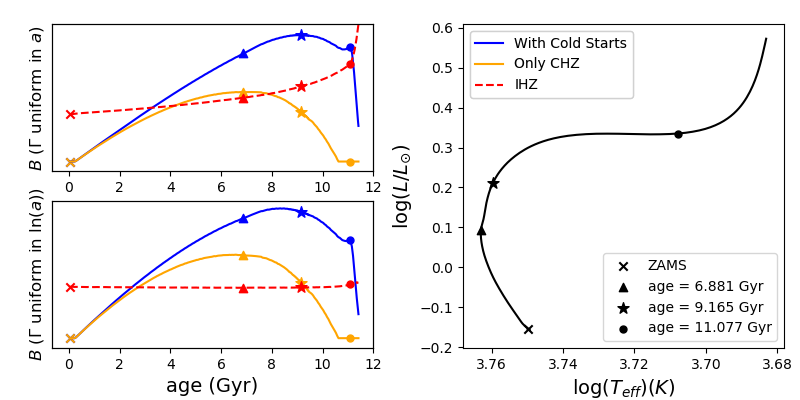}
    \caption{{\it Left:} Relative biosignature yields, $B$, of a 1 $M_\odot$ star as a function of age under two assumptions for $\Gamma$, the distribution of exoplanets in orbital distance (upper and lower subplots). $B$ is a strong function of both time and habitability model. The colored curves reflect three habitability models: that all planets in the instantaneous habitable zone (IHZ) are equally good targets (red), and that planets have uniform probability per unit time of generating biosignatures if they are in the continuous habitable zone (CHZ, orange) or the IHZ (i.e.\ assuming that ``Cold Start'' planets can thaw, blue). The red curve has arbitrary vertical scaling with respect to the other two. The choice of habitability model has a strong effect on the relative merit of stars of different ages. The positions of the maxima of the $B$ curves for the case where $\Gamma$ is uniform in $a$ are marked in each subplot and on the HR diagram of the evolution of the star ({\it Right}).}
    \label{B_evolution}
\end{figure}

\subsection{Time evolution}
\label{time_evolution}

As a first step for comparing our metrics for relative biosignature yields between stars, we would like to know how they vary as a function of time when all other stellar properties are held constant. We shall use the model of a sun-like star used in Figures \ref{Sun_hz} -- \ref{past_hab_lifetime_age} and calculate our metrics for it as it evolves in time. For each time step in the model evaluation, we shall calculate the different $B$ values using the evolutionary history of the star's luminosity and effective temperature. We continue the star's evolution beyond the subgiant phase, up until shortly before the helium flash.

In Figure \ref{past_hab_lifetime_age}, we illustrated how the habitable duration, $\tau$, changed as a function of time. This helps to inform how we would expect our metrics to vary for stars of different ages. 
If one uses a form of $\Gamma$ \added{that is} uniform in $a$, then the metrics that use $H(a,t)$ that is linear in $\tau$  would be directly proportional to the area under the curve for $\tau$. One can see that Figure \ref{past_hab_lifetime_age} could be used to infer how the metric with $\Gamma$ uniform in $a$ and $H$ including Cold Start planets changes in time. It would appear that, in the age regime that the figure covers, the area under the curve is continuously increasing. If one only considers the plateau regions of the $\tau$ curve representing the CHZ, then it is less clear how the area, and hence the value for $B$ would change in time, as the region under the curve would get taller, but narrower.

We graph the evolution of our metrics in time in Figure \ref{B_evolution}. The left panels shows the time evolution of metrics with 3 different assumptions for $H(a,t)$, and the 2 assumptions for $\Gamma$ in the upper and lower subplots. The right panel shows the star's evolution on the HR diagram starting on the zero age main sequence (ZAMS). The different maxima of the $B$ curves for the $\Gamma$ uniform in $a$ case are marked on the evolutionary track and the $B$ evolution curves. One can observe that the blue and orange curves, representing formulations of $H(a,t)$ that are dependent on the habitable duration, appear very similar for both assumptions of $\Gamma$, though those for the uniform in $\ln(a)$ case appear to peak a bit earlier.
Looking at the blue curves, one can see the metrics which consider Cold Start planets to be habitable continue to increase until they peak for older stars near the end of the main sequence. Interestingly, they appear to have secondary maxima on the subgiant branch, which likely corresponds to the onset of degeneracy in the core. After this point, the star's luminosity increases quickly and regions that enter the habitable zone don't remain habitable for long, causing B to drop off sharply.

The metrics which only consider the CHZ in $H$ behave similarly to the Cold Start metrics, but peak around 2 Gyr earlier. It still appears that older stars are preferred by these metrics, but not quite as old as those of the Cold Start metrics. As the star reaches the subgiant branch, there comes a point where the inner edge of the CHZ recedes to an extent where none of the planets in the current habitable zone have been continuously habitable  since their formation. The CHZ metric goes to zero here and remains there. 

For treatments of $H$ that do not depend on the habitable duration, but only depend on the IHZ, the time evolution is much different. Between the upper and lower left subplots there is a substantial difference between the IHZ metrics depending on assumptions for $\Gamma$. The metric with the instantaneous $H$ and $\Gamma$ uniform in $a$ depends primarily on the width of the habitable zone, so it roughly traces the luminosity evolution of the star. As the star gets gradually brighter on the main sequence the metric value also increases gradually. When it reaches the subgiant branch and the luminosity of the star increases dramatically, so too does the IHZ metric. Given that this metric is primarily related to the luminosity of the host star, this would imply that giant stars are the best candidates to detect biosignatures around. Since we know that giant evolved stars pose major obstacles to habitability, this suggests that this metric is a poor proxy for habitability. On the other hand, the metric for instantaneous habitability with $\Gamma$ uniform in $\ln(a)$ is almost constant in time. This tells us that this metric is insensitive to the age of the target stars, which may also limit its usefulness in a search for biosignatures. In the following section, we shall see that not only is this metric insensitive to stellar ages, it is also barely dependent on stellar masses. This severely limits the utility of such a metric, as it gives similar values for very different stars.

\begin{figure}
	\centering
	\includegraphics{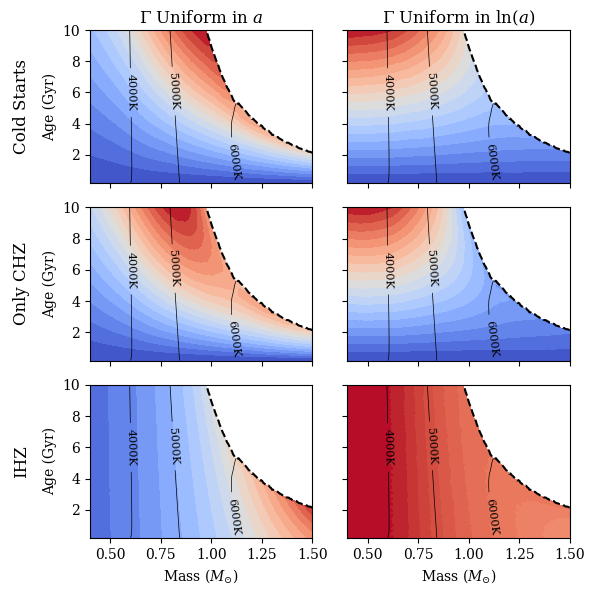}
	\caption{Different assumptions about the distribution of planets in orbital distance, $\Gamma$, and the emergence of biosignatures, $H(a,t)$, can have a major effect on which stellar populations are preferred in a search for biosignatures. For instance, the assumptions represented in the lower right panel favor K dwarfs of any age as targets for biosignature searches, while the assumptions in the upper left panel favor old G dwarfs.
	Columns correspond to $\Gamma$, the frequency of terrestrial planets, and rows to different models of biosignature genesis. ``Cold Starts'' refers to models in which planets can ``thaw out'' upon entering the habitable zone (see Section \ref{cold_starts_section}) while ``Only CHZ'' refers to planets that have spent their entire histories in the continuously habitable zone. IHZ refers to planets in the present day (instantaneous) habitable zone. The dashed lines are the Terminal Age Main Sequence. The color scale is set to blue at zero and red at the maximum biosignature yield for each subplot; as such, colors shouldn't be compared between subplots. Also included are contours of constant stellar effective temperature.}
    \label{large_grid_compare}
\end{figure}

\subsection{Mass and Age Dependence}
\label{mass_age_dep}
We will next observe how our biosignature yield metrics vary for a wide variety of stars. The key variables that affect the values of our metrics are the stellar mass and age, so we modelled stars with a range of masses and ages to see how their relative biosignature yields, $B$, varied. We used a simple stellar model in MESA \citep{MESAPaper2011,MESAPaper2013,MESAPaper2015,MESAPaper2018,MESAPaper2019}, described in Appendix \ref{Stellar_models}, and varied the stellar masses and ages, holding the metallicity and helium mass fractions constant at roughly solar values. Using the outputs from these MESA models, we calculated values for the expected biosignature yield for different assumptions about $\Gamma$ and $H(a,t)$.

Figure \ref{large_grid_compare} demonstrates how our values for the metrics vary over a grid of masses and ages. For these subplots, the columns represent assumptions about the form of $\Gamma$, the distribution of exoplanets. The rows indicate assumptions about the functional form of $H(a,t)$ the probability of a planet hosting biosignatures at a given semimajor axis. Given the differences in colormaps between the different subplots, it is clear that different assumptions can greatly affect whether a star of given mass and age is a good candidate to host biosignatures. 
Recall that because we are studying {\it relative} biosignature yields, we can only compare between stars that share a set of assumptions about $H$ and $\Gamma$, and that we have assumed that $\Gamma$ is independent of stellar mass and age. The color scales in Figure \ref{large_grid_compare} should not be compared between the different subplots. The color scale on each of the subplots goes from zero in blue to the maximum value of metric in red.


Various assumptions about $\Gamma$ and $H$ result in different locations for the maxima of the metrics, $B$. If one considers Cold Start planets to be habitable and assumes that $\Gamma$ is uniform in $a$, then the maximum values of the metric hug the terminal age main sequence (dashed black line in Fig \ref{large_grid_compare}), with a maximum corresponding to older G stars. If we keep the assumption that Cold Starts can be habitable and change $\Gamma$ to be uniform in $\ln(a)$, then the maximum values of the metric are for low mass stars with large ages. Choosing $H$ to only consider planets in the CHZ slightly modifies the shape of the colormaps for both assumptions about $\Gamma$. In the CHZ only, $\Gamma$ uniform in $a$ case, the highest $B$ values still seem to be close to the terminal age main sequence, but the maximum is towards lower stellar masses than the Cold Starts case, leaning towards older late-G or early-K stars. For the CHZ only, $\Gamma$ uniform in $\ln(a)$ case, the highest metric values are a bit more focused towards lower mass older age values than in the Cold Starts, $\Gamma$ uniform in $\ln(a)$ case. This preference towards lower stellar masses for the Only CHZ $H$ formulation is expected, as lower mass stars evolve less quickly on the main sequence, resulting in a lower proportion of Cold Start planets.

The metrics that consider only the IHZ of a star at a given age vary dramatically from the other assumptions about $H$, as they don't give any consideration to the evolutionary history of the stars, and only consider the current location of their habitable zones. Using the IHZ $H$ and $\Gamma$ uniform in $a$, the maximum value of the metric is towards high mass stars near the end of the main sequence. This makes sense as this metric favors habitable zone width and more luminous stars would have wide habitable zones. The IHZ, $\Gamma$ constant in $\ln(a)$ metric is fascinating in that it does not vary as much as the other metrics. As such, one can see that the colormap seems much redder than the other plots, with a maximum towards lower stellar masses, and not much age dependence.

\section{Application to Specific Target Stars}
\label{ranking_targets}
\begin{figure}
    \centering
    \includegraphics{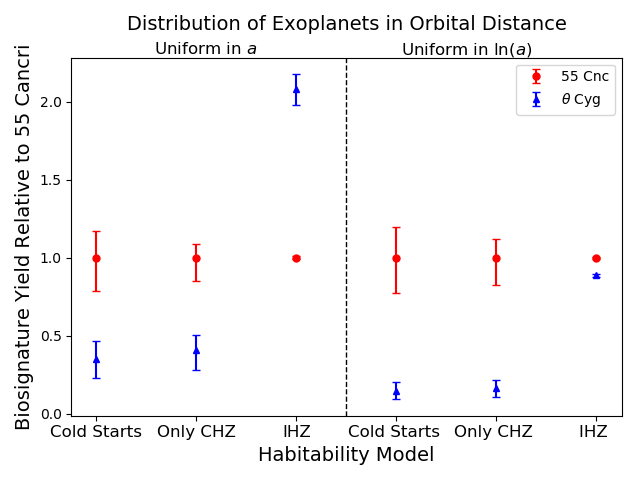}
    \caption{Comparison of expected biosignature yields, $B$, for 55 Cancri and $\uptheta$ Cygni. The relative values of the metrics are compared for the six different models of habitability and exoplanet distribution with orbital distance. ``Cold Starts'' refers to models in which planets can ``thaw out'' upon entering the habitable zone (see Section \ref{cold_starts_section}) while ``Only CHZ'' refers to planets that have spent their entire histories in the habitable zone. IHZ refers to planets in the present day (instantaneous) habitable zone. A description of the models used to generate these values and their uncertainties is in the appendices. }
    \label{metric_value_compare}
\end{figure}

The relative biosignature yield is meant to be compared between potential direct imaging target stars in order to determine which are more likely to host biosignatures. 
In future studies, we aim to characterize the biosignature yields for an entire list of target stars for future direct imaging missions. As a first step towards this goal, we perform our analysis for two well studied stars, and assess the extent to which different assumptions about $\Gamma$ and $H$ affect the relative values of $B$.

In Appendix \ref{example_systems}, we describe how we have calculated $B$ for two example stars: 55 Cancri and $\uptheta$ Cygni. 55 Cancri is an old K dwarf and we would expect it to be a better direct imaging candidate in terms of expected biosignature yield than $\uptheta$ Cygni, a comparatively younger F star. Note that, while 55 Cancri is a well known planet host star, here we treat it as if it doesn't host known planets, and instead calculate biosignature yields based on estimates of the planet occurrence rate. Here 55 Cancri acts as a fiducial stand-in for an example  K star with well characterized properties. We fit stellar models to both of these stars and calculated the metric values for the optima and the regions of parameter space around them. For model fits of both stars, we assessed the extent to which metric values varied within the range of uncertainty in observed stellar properties.

In Figure \ref{metric_value_compare}, we compare the values of the metrics for the best fit models of the two stars. 
Approximate uncertainties for the metric values are given via the spread in values of the metric within the 1-sigma contour for $\chi^2$. As the absolute values of the metrics are irrelevant, given that we don't know the normalization of $\Gamma$ or the probability of biosignatures emerging per unit time, we report values of the metrics relative to those of the optimum of 55 Cancri. We plot the 6 metrics sorted by assumptions for $\Gamma$ (columns on the top) and assumptions for $H(a,t)$ (on the bottom). 

For formulations that depend on habitable duration, we can see that 55 Cancri is favored over $\uptheta$ Cygni, as expected. The stars have sufficiently distinct masses and ages that the details of Cold Starts or the functional form of $\Gamma$ does not change their relative merit\replaced{, however}{. However, }it is clear that finer gradations and the sizes of our error bars will depend on such choices. 


The IHZ formulation of $H(a,t)$ has the largest difference from the other metrics, and indeed the relative merits of the two stars \replaced{changes}{change} dramatically depending on the true form of $\Gamma$. One can observe that if $\Gamma$ is uniform in $a$, and $H(a,t)$ depends only on the IHZ, then $\uptheta$ Cygni actually scores significantly higher than 55 Cancri. 
This makes sense because under this assumption $B$ depends primarily on habitable zone width, so a more massive star such as $\uptheta$ Cygni would have a much wider habitable zone in linear space. Similar to as we saw in the in previous sections, for the IHZ metric with $\Gamma$ uniform in $\ln(a)$, the two stars rank very similarly due to the general insensitivity of this metric to mass and age.
These metrics have the virtue of very small error bars because they depend only on the luminosity and temperature of the stars, which are very well known, and not on their habitable histories, which are not. 
This appears to be offset, however, by the fact that the IHZ metrics differ so much from the other, more physically motivated metrics, making the IHZ a poor approximation for studies of habitability and biosignature yield.

For well studied stars, such as 55 Cancri and $\uptheta$ Cygni, it is straightforward to distinguish which star ranks higher according to our metrics. As seen in Figure \ref{metric_value_compare}, the uncertainties are small enough that there is no overlap between them, but for stars with less precisely measured properties, it may be more difficult to classify which have higher relative biosignature yields. \added{In general, many prospective direct imaging targets will be very bright and nearby, and so will be similarly well characterized to 55 Cnc and $\uptheta$ Cyg.  Others may have stellar properties that are less well determined, or may have independent measurements of other stellar properties such as asteroseismic modes and model independent age constraints.}

\added{Given the sensitivity of biosignature yields to a star's habitable history, it is important to understand the sensitivity of $B$ to the uncertainty in stellar ages and other observables used to constrain stellar evolutionary tracks. As discussed in Section \ref{age_uncertainty}, age measurements are often determined by fitting stellar models to observed stellar properties. Prior age measurements are not required to fit a stellar evolutionary track to a star, but they can be used as further constraints for model fitting. 
While we plan to perform a full, statistically rigorous sensitivity analysis in future work, here we will provide a preliminary analysis of the sensitivity of our metrics to observational constraints on age.

As a first pass to determine the age sensitivity of $B$, we consider the case where we vary only the ages of stellar models and see how the metrics change in response. Using the data used to generate Figure \ref{B_evolution} we observe how $B$ changes when all parameters other than age are held constant. Around the current age of the sun (roughly 4.57 Gyr), we varied the age by 15\%, a comparable uncertainty to what one would expect in the case of gyrochronology age measurements. With this uncertainty in age, metrics varied by as much as 14.4\% for the cold starts, $\Gamma$ uniform in $a$ case, to only 8.3\% on average for the CHZ, $\Gamma$ uniform in $\ln(a)$ case.
Looking at the curves in Figure \ref{B_evolution}, one can see that the variability of the metrics depends on the the region of parameter space an age measurement falls in. In future studies, it may be informative to investigate which metrics for biosignature yield are most affected by stellar age uncertainty.

We next considered the case where independent age constraints can be used to improve the accuracy of a model fit. For the two example stars, 55 Cnc and $\uptheta$ Cyg, we saw how $B$ varied in the case of realistic uncertainties in observable properties in Figure \ref{metric_value_compare}. We next add age as an additional constraint in our $\chi^2$ formulation. We used an artificial age constraint, setting the true value of age to be the fit optimum and allowing it to vary by 15\%. To illustrate how adding a constraint in stellar age affects the uncertainty of our metrics, we will focus on the variability of one particular metric: the formulation of $B$ which only considers the CHZ and assumes $\Gamma$ is uniform in $\ln(a)$. When no age constraints are applied, this metric has an average uncertainty of 14.6\% for 55 Cnc and 32.3\% for $\uptheta$ Cyg. When we include an age constraint with an uncertainty of 15\%, the uncertainties in $B$ decrease to 11.7\% and 22.6\% for 55 Cnc and $\uptheta$ Cyg respectively. This demonstrates that adding additional age constraints such as those from gyrochronology can greatly improve the precision of our metrics for biosignature yield. Even without age constraints, we can specify the values of our metrics to a precision that would be useful to statistically characterize which stars have the highest probability of hosting biosignatures. With additional age constraints, or potentially more precise measurements of spectroscopic quantities, one would be able to use our metrics to unambiguously rank the individual stars in a target list.}
\deleted{More precise measurement of stellar properties may be required in order to more tightly constrain stellar masses, ages, and evolutionary tracks.}

\section{Discussion and Future Work}

We have developed a framework for comparing the predicted biosignature yields for target stars of future direct imaging missions. In this work,
we have shown how biosignature yield depends sensitively on knowledge of precise stellar properties of potential target stars (see Appendix \ref{example_systems}), on the occurrence rate of terrestrial planets in orbital separation, and, most crucially, on what assumptions are made about the formation of biosignatures (Section~\ref{B_formulation} and Figure~\ref{metric_value_compare}).  
We have constructed a metric for {\it relative} biosignature yield between stars, $B$, and shown how it can be calculated from the planet occurrence rate $\Gamma$ and a model of biosignature formation, given as $H(a,t)$, the probability that biosignatures have arisen on a planet at orbital separation $a$ around a star of age $t$. 

Studies that use only the present-day Habitable Zone as a proxy for habitability will implicitly assume that planets of any age and evolutionary history have an equal chance of supporting biosignatures so long as they are in the \textit{instantaneous} habitable zone, which is both inherently unlikely and at odds with the history of life on Earth. Metrics for biosignature yield that only consider the instantaneous habitable zone lead to misleading conclusions about which populations of stars would have the highest biosignature yields. For the case where $\Gamma$ is uniform in $a$, high mass stars and older stars with higher luminosities are preferred due solely to their wider habitable zones, ignoring the major obstacles to habitability posed by such stars. When $\Gamma$ is instead uniform in $\ln(a)$, one finds that the biosignature yield is very insensitive to the mass or age of a star, significantly reducing the utility of such a metric.

We demonstrated how relative biosignature yield depends on stellar mass and age in Section \ref{B_variability}. In general, if one assumes that biosignatures depend on a planet's habitable duration, then older stars and stars with slower luminosity evolution are favored because they give their planets the longest time to develop biosignatures. This favors the more numerous and longer-lived K and G dwarfs, with closer but smaller habitable zones, over  more luminous but rarer F stars. The details of whether K dwarfs are favored over G dwarfs depends on the underlying distribution of terrestrial planets, and on whether Cold Start planets are habitable. This has important implications for direct imaging mission design because it affects the angular separation and spectral energy distribution of typical target stars. 

This work is meant to introduce a formalism for comparing the biosignature yields around various target stars for direct imaging missions. As a first step, we fit stellar models to two well-studied stars, 55 Cancri and $\uptheta$ Cygni, and saw how they compared in regards to various metrics for biosignature yield. Additionally, \replaced{in Appendix \ref{example_systems}}{in Section \ref{ranking_targets}} we investigated how uncertainties in the relative biosignature yields of these stars depends on knowledge of their stellar parameters, \replaced{particularly their mass and age}{particularly in the case when model independent age constraints are available}. We find that their relative merits are rather insensitive to the assumption of whether Cold Start planets host biosignatures, but very sensitive to whether one uses the present-day Habitable Zone or a more realistic model.

\subsection{Future Work}
The clear next steps going forward would be to apply the methods introduced in this paper to an entire stellar population, such as the list of target stars for future direct imaging missions. Such target lists have not been finalized, but preliminary lists of potential nearby direct imaging target stars are available. In calculating the total biosignature yield for a direct imaging mission, one could observe which samples of target stars are most likely to host biosignatures.
One of the main difficulties for calculating the biosignatures yields for a sample of target stars would be that many \deleted{nearby} stars \added{may} lack sufficiently precise uncertainties in their observed properties as well as derived properties such as their masses and ages. To calculate biosignature yields for such a sample of stars would require first obtaining accurate masses and ages for each of the target stars.
For the two example target stars we modelled in this paper, we obtained masses, ages, stellar evolutionary histories, and biosignature yields by fitting stellar models to observed properties using an optimizer. Despite the fact that these example stars were specifically chosen because they had well characterized stellar properties, it took hundreds of iterations for an optimizer to find a suitable model fit for each of the stars. Given that each model evaluation took around 5 minutes on the 16 core CPU used in this study, it becomes clear that applying the same methods to hundreds of target stars in a catalog would be very computationally intensive. To calculate the biosignature yields for a sample of target stars, we would either need to use a supercomputer or find a more computationally efficient means to obtain model fits. 

Given the challenges involved in fitting models for an actual stellar population, it would be significantly easier, though potentially less informative, to instead calculate biosignature yields for a synthetic stellar population. In that case, the masses and ages of the stars would be known \textit{a priori}, and one would only need to run one stellar model per star. 

While for the purposes of this paper we only used a rough estimate for the uncertainties in $B$, in future studies we will take a more statistically robust approach. To deal with the computationally intensive nature of model fitting, we will explore solutions such as Gaussian process emulation of MESA model outputs. Alternatively, methods such as isochrone fitting would also greatly reduce the computation time, but would reduce the precision of the model fits and wouldn't take into account all of the free model parameters and outputs. With a less computationally intensive function constructed to simulate the model outputs, it becomes feasible to use Markov Chain Monte Carlo (MCMC) methods to obtain the best model fits and uncertainties in the fit parameters.

Future work is also needed to address several additional important points. Potential direct imaging target stars have a heterogeneous and heteroskedastic suite of measured fundamental properties, including abundance patterns, asteroseismic measurements, \added{gyrochronological age constraints, }    interferometrically measured angular radii, spectral energy distributions, and parallaxes. Assessing their habitable zone histories requires incorporating this information into stellar evolution models in a consistent way. Stellar evolution models, such as MESA, are themselves subject to uncertainty, such as from incomplete treatments of rotation, convection and diffusion, and uncertainties in opacities and equations of state. These uncertainties should be explored and incorporated into uncertainties in biosignature yield. In future studies, we plan to conduct a sensitivity analysis on biosignature yields with respect to both uncertainties in observed stellar properties as well as inherent stellar modelling uncertainties.

The habitability of Cold Start planets has received scant attention, but since these planets likely compose a large fraction of all discovered exoplanets in the IHZ, they warrant further study. Such planets not only add to the total potential biosignature yield, but they are found in the outer parts of habitable zones, so excluding them has important implications for the necessary inner working angle of direct imaging missions. 

The relative biosignature yield of stars also depends sensitively on the \replaced{slope}{dependence} of $\Gamma$ \replaced{with}{on} $a$, which is poorly constrained at the radii and orbital separations where most direct imaging targets will be found.  Its dependence on stellar properties such as mass, metallicity and effective temperature is also poorly known, but could have major implications for the relative merits of K versus G stars. Future studies should investigate the extent to which different functional forms of the dependence of $\Gamma$ on these properties affects the relative biosignature yields.

\subsection{Overview}
In this paper, we have introduced a metric for the relative yield of biosignatures for target stars of future direct imaging missions. We have seen how these metrics vary for different types of stars and outlined our plans for future studies using the biosignature yield formalism defined here. This work is focused on paving the way for future research and, as such, is not overly focused on results. That being said, through modelling the biosignature yields of stars, we have gained insight into the necessities for direct imaging mission design. We find that robust planning for future missions such as HabEx and LUVOIR, including discussions of their relative merits, will require:
\begin{itemize}
\item An explicit model of biosignature appearance that incorporates the history of habitability of planets
\item An understanding of the habitability of Cold Start planets
\item Knowledge of the terrestrial planet occurrence rate with orbital separation and stellar mass
\item Precisely known stellar properties of all potential target stars, especially mass and age
\item Robust and consistent reconstructions of the evolutionary histories of all potential target stars
\end{itemize}

\acknowledgments
{NWT and JTW acknowledge James F.~Kasting for his feedback and guidance on this project. NWT acknowledges Marc Pinsonneault and the participants of the 2019 MESA Summer School for their assistance with stellar modelling in MESA. 

The Center for Exoplanets and Habitable Worlds and the Penn State Extraterrestrial Intelligence Center are supported by the Pennsylvania State University and the Eberly College of Science.

This work has made use of data from the European Space Agency (ESA) mission
{\it Gaia} (\url{https://www.cosmos.esa.int/gaia}), processed by the {\it Gaia}
Data Processing and Analysis Consortium (DPAC,
\url{https://www.cosmos.esa.int/web/gaia/dpac/consortium}). Funding for the DPAC
has been provided by national institutions, in particular the institutions
participating in the {\it Gaia} Multilateral Agreement.

This research has made use of NASA's Astrophysics Data System Bibliographic Services. 
}

\appendix

\section{Stellar Modelling}
\label{Stellar_models}

In this study, we ran a variety of stellar models using the MESA stellar evolution code \citep{MESAPaper2011,MESAPaper2013,MESAPaper2015,MESAPaper2018,MESAPaper2019}.
In Section \ref{mass_age_dep}, where we model stars for a wide variety of masses and ages, we used a simple stellar model in MESA. For a specified stellar mass and age, this model evolves the star from the premain sequence up until the specified age. If the age given is too large, this model will instead end on the Terminal Age Main Sequence (TAMS), defined via core hydrogen depletion (i.e. when the center radial cell of the model has a hydrogen mass fraction of 0.001 or less).
Besides mass and age, this simple model holds all input parameters constant, with initial helium and metal mass fractions of $Y_i = 0.28$ and $Z_i=0.02$ respectively. With this framework for simple stellar models, we calculated models for a large grid of masses ranging from 0.4 - 1.5 $M_\odot$ and ages ranging from 0.2 - 10.0 Gyr.

The models we used in Sections \ref{metrics_section} and \ref{time_evolution}  were similar to the simple MESA model used for the larger mass age grid, but they incorporated more sophisticated input physics such as element diffusion, convective overshoot, and more realistic boundary conditions.
We used convection and convective overshoot terms from an optimized Solar Model and held them constant for all models. In Appendix \ref{example_systems}, where we calculate the values of the biosignature yield metrics for example stars, we use this model for 55 Cancri, fitting for the observed stellar properties.

We used a slight variation of the previously mentioned stellar model for the example star $\uptheta$ Cygni. The main difference between our stellar models for $\uptheta$ Cygni and that of the other stars was that we didn't include the effects of element diffusion. This is due to a known stellar modelling problem with F stars. Because of their very thin outer convective zones, stellar models for F stars which include gravitation settling of elements have the problem that the photosphere quickly depletes most of its heavier elements and helium \citep{Guzik2016}. Since we do not observe this extreme depletion of heavy elements in the spectra of F stars, there must be some physical process that opposes gravitational settling which stellar models don't take into account.

\section{Example systems}
\label{example_systems}
\subsection{55 Cancri A}
\label{55Cnc_section}
\begin{figure}
    \centering
    \includegraphics{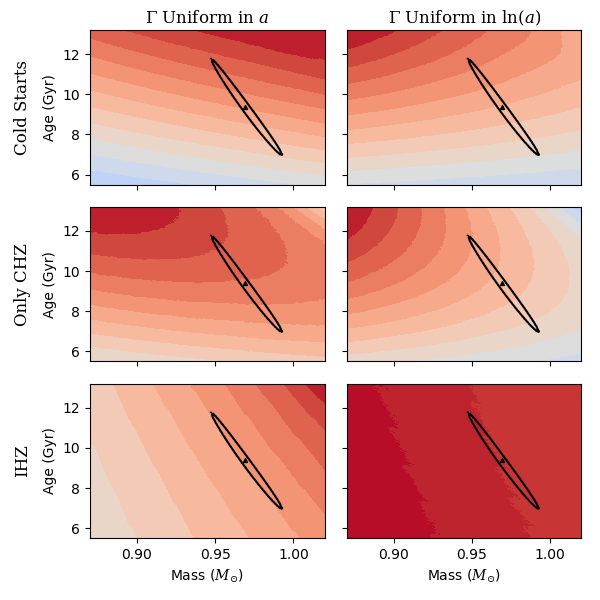}
    \caption{Expected biosignature yield metrics $B$ as a function of stellar mass and age for the 55 Cancri model. Columns of subplots correspond to different assumptions about the functional form of $\Gamma$, the distribution of rocky exoplanets in orbital distance, while rows are for different assumptions about $H(a,t)$, the probability that a planet at a given distance and age hosts biosignatures. Metrics were computed using a calibrated MESA stellar model for 55 Cancri, holding all other properties besides mass and age constant. Also included is the 1-sigma contour of mass and age values for the model fit. Color scales are for relative values with the maximum value of a metric in red and zero in blue. As such, they shouldn't be compared between subplots.}
    \label{55Cnc_compare}
\end{figure}

Since these metrics are meant to be compared between stars to compare their relative biosignature yields, we tested our metrics on example stars and saw how the metrics varied  within the range of measurement uncertainties and how they compared to each other. The first of these example stars was 55 Cancri. 

55 Cancri (HD 75732) is a well studied star due to the fact that it hosts 5 exoplanets discovered via the radial velocity technique \citep{vonBraun2011}. Because one requires precise stellar properties to accurately determine the properties of exoplanets, it has been the target of follow up studies to constrain its properties such as mass, age, luminosity, and metallicity. As such, the measured properties of this star are more precise than most target stars for direct imaging missions. 55 Cancri is peculiar among potential target stars in that it has a very high metallicity, with measurements of $[\text{Fe/H}]= 0.31\pm0.04$\citep{Marcy2002,ValentiFischer2005}. The spectral type for 55 Cancri ranges between G8V \citep{Ligi2016} to K0 IV-V \citep{Gray2003} in the literature.
With obtaining a model fit to 55 Cancri, there is a major degeneracy between mass and age, so uncertainties are higher than one might expect. Mass estimates for 55 Cancri range from  $0.874\pm0.013 M_\odot$ to $1.015 \pm 0.051 M_\odot$ \citep{Ligi2016,Crida2018}. The age of 55 Cancri is similarly uncertain with values ranging from $10.2\pm 2.5$ to $13.19\pm 1.18$ Gyr \citep{vonBraun2011,Ligi2016}. However, estimates of the age of 55 Cancri from stellar rotation give ages in the range of 7.4 to 8.7 Gyr, inconsistent with most model fits \cite{Mamajek2008}.

For our model of 55 Cancri we calibrated using target values for $\log(L)$ and $\log(R)$ from \citet{vonBraun2011} and $[\text{Fe/H}]$ values from \citet{ValentiFischer2005}. Since we now have updated parallax measurements from {\it Gaia}, we updated the target luminosity and interferometric radius values to reflect the new distance. {\it Gaia} reports a parallax measurement of $79.427 \pm 0.078$ mas for 55 Cancri, resulting in a distance of $12.590\pm 0.012$ pc \citep{GAIArelease2}. We used the angular diameter and bolometric flux from \citet{vonBraun2011} to calculate $R=0.9621\pm0.0055 R_\odot$ and $L= 0.6058\pm0.0088 L_\odot$.
Varying the initial helium mass fraction, initial [Fe/H], Mass, and age of the star, we fit a model to the observed log(R), log(L) and [Fe/H] of 55 Cancri via optimization with the Nelder-Mead simplex algorithm. We optimized using a $\chi^2$ formulation of $\chi^2 = \sum_{i=1}^N \frac{(O_i -E_i)^2}{  \sigma_i^2}$, with N Observed output values and expected target values as $O_i$ and $E_i$ respectively, and uncertainties as $\sigma_i$. We find an optimal fit for 55 Cancri with input parameters of $Y_i=0.2680$, $[\text{Fe/H}]_i=0.3720$, $M=0.9690 M_\odot$, and age = 9.389 Gyr, which results in a stellar model that matches the observables almost exactly.
The values we obtain for the mass and age of 55 Cancri are consistent with the literature values, despite the fact that we used a different stellar model.

While 55 Cancri is a planet host star, for this study we will treat it as if it were a well characterized star without known planets. Therefore, to calculate expected biosignature yields we use the formulation of $B$ in Section \ref{B_formulation}. We  ran a large grid of models around the $\chi^2$ optimum, holding initial $Y$ and [Fe/H] constant, and only varying mass and age. For each model, we calculated the values of our metrics and observed how they varied within the uncertainty for mass and age. Figure \ref{55Cnc_compare} shows the relative biosignature yields given by each metric. It is immediately apparent that, compared to Figure \ref{large_grid_compare}, Figure \ref{55Cnc_compare} is much redder, indicating that there is less variability in the metric values than there was over the larger mass and age range. Again we use a color scale where red corresponds to the maximum value of each metric in the mass and age range, and blue corresponds to zero. To gain a sense of the uncertainty in the mass and age for the model, we plotted what we refer to as a ``1-sigma" contour on each of the subplots. We used a heuristic argument to approximate a 1-sigma level from our $\chi^2$ values, since when $\chi^2 = N$, the number of target values fit, the average separation between model outputs and target values will be about 1 sigma. Looking at the 1-sigma contours in mass and age we can see how correlated mass and age are, and the difficulties involved in breaking the degeneracy.

The amount of variability of the metrics within the 1-sigma contour varies a lot between different formulations. In general the metrics with a uniform $H(a,t)$ over the IHZ have much less variability than the metrics which use an $H(a,t)$ that is proportional to the habitable duration. One can see that some of these metrics appear to have a substantial range in values within the uncertainties of the model fit, but to determine the usefulness of the metrics we need to see how they compare between stars, and whether the uncertainty in metric values will affect which stars are favored in terms of relative biosignature yields (see Section \ref{ranking_targets}).


\subsection{Theta Cygni}

\begin{figure}
    \centering
    \includegraphics{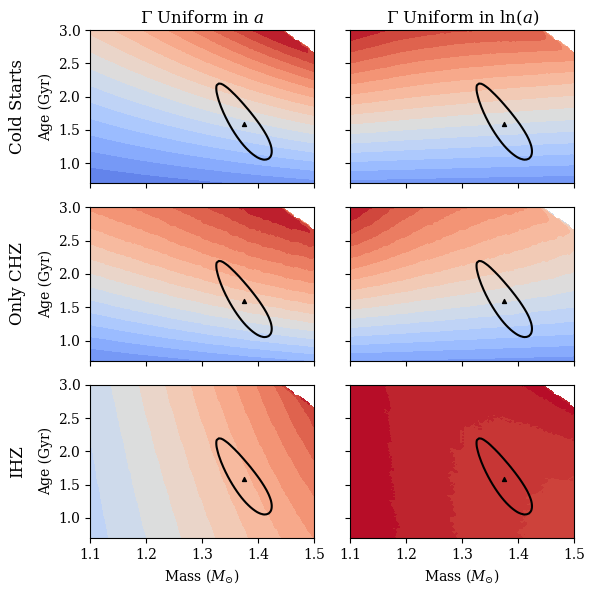}
    \caption{Expected biosignature yield metrics $B$ as a function of stellar mass and age for the $\uptheta$ Cygni model. Columns of subplots correspond to different assumptions about the functional form of $\Gamma$, the distribution of rocky exoplanets in orbital distance, while rows are for different assumptions about $H(a,t)$, the probability that a planet at a given distance and age hosts biosignatures. Metrics were computed using a calibrated MESA stellar model for $\uptheta$ Cygni, holding all other properties besides mass and age constant. Also included is the 1-sigma contour of mass and age values for the model fit. Color scales are for relative values with the maximum value of a metric in red and zero in blue. As such, they shouldn't be compared between subplots.}
    \label{ThetaCyg_compare}
\end{figure}

We then turned our attention to another potential target star for direct imaging missions, $\uptheta$ Cygni. This is a nearby F star and, because of the shorter main sequence lifetime and faster evolution, we anticipate that it will score worse for most of the metrics for long term habitability. $\uptheta$ Cygni A (HD 185395) was the brightest star observed by the Kepler mission, and unlike many other potential target stars, it has high precision asteroseismic measurements of stellar oscillations. The high quality of its asteroseismic data has led $\uptheta$ Cygni to be the target of past studies such as that of \citet{Guzik2016}. They characterize it as a F3 star with a mass between 1.35 - 1.39 $M_\odot$  and an age between 1.0 - 1.6 Gyr. 

To fit a stellar model to $\uptheta$ Cygni we used target values taken from \citet{Guzik2016}. While $\uptheta$ Cygni has high quality measurements of stellar oscillation frequencies, we did not attempt to fit the known mode frequencies with MESA. Because most target stars for direct imaging missions do not have asteroseismic measurements of comparable accuracy, we fit a stellar model to $\uptheta$ Cygni as if it were a such target star without such high quality measurements. This approach also allowed us to see whether values for stellar masses and ages obtained without asteroseismic data are consistent with those using stellar oscillation data.

From the spectral analysis of \citet{Guzik2016}, we have target values of $T_{\rm eff}=6697 \pm 78$ K, $[\text{Fe/H}] = -0.02 \pm 0.06$, and $\log(g)= 4.23 \pm 0.03$ in cgs. Similarly, we used their stellar radius value to get a target value of $\log(R)=0.173 \pm 0.009$ in solar units. It should be noted that since $\uptheta$ Cygni is such a bright star and is part of a multiple star system, the newer {\it Gaia} distance measurements for it aren't actually more precise than the earlier Hipparcos measurements. Therefore we used the target values as presented in \citet{Guzik2016} and didn't adjust for newer distance measurements.

We used a stellar model for $\uptheta$ Cygni described in Appendix \ref{Stellar_models}, which was the same as that of 55 Cancri, but didn't include element diffusion. To optimize our stellar model we used the same methods for optimization as we did for 55 Cancri, but we used different target values in our $\chi^2$. We calibrated our model fitting to the target values of log(R), $T_{\rm eff}$, surface [Fe/H], and log(g). We found that an optimum fit was achieved when $Y_i =0.2564$, $[\text{Fe/H}]_i = -1.015 * 10^{-3}$, $M=1.376 M_\odot$, and age= 1.588 Gyr, yielding a very good fit to the target values.
We note that our optimum values for the mass and age of $\uptheta$ Cygni are consistent  with the literature values from \citet{Guzik2016}. 

We then ran a grid of models in mass and age around the optimum and plotted the colormaps. One can clearly observe in Figure \ref{ThetaCyg_compare} that the colormaps for the region of parameter space around the optimum for $\uptheta$ Cygni have a very different shape than those for 55 Cancri, in some cases having maxima in different locations. 
Particularly, for the Only CHZ assumption for $H(a,t)$, the maximum values of the metrics are in a different region of parameter space than those of 55 Cancri if one uses the assumption that $\Gamma$ is uniform in $a$. Whereas the 55 Cancri colormap had a maximum towards low mass, older stars, the $\uptheta$ Cygni values for this metric prefer older high mass stars.



On Figure \ref{ThetaCyg_compare} we also plot the 1-sigma contours for the optimum model fits. As with 55 Cancri we use the term 1-sigma to refer to the contour where $\chi^2$ equals the number of observed target values fit to. While we had three for the case of 55 Cancri, here we have 4 target values, the log(R), $T_{\rm eff}$, [Fe/H], and log(g). It is interesting to note that the 1-sigma contour here seems less highly correlated between mass and age. Perhaps this is due to including log(g) in the $\chi^2$. One can see that the values of the metrics within the 1-sigma contours are pretty variable, but since we have different color scales for both figures it is difficult to compare their ranges of values by eye. Therefore, the relative values of the metrics are plotted in Figure \ref{metric_value_compare}.

\bibliographystyle{aasjournal}
\bibliography{sources}
\end{document}